\documentclass[twocolumn]{aastex631}

\usepackage{hyperref}
\usepackage{verbatim}
\accepted{July 4, 2022}
\usepackage{amsmath}
\usepackage{amsfonts}
\shorttitle{PSR J0952$-$0607's High Mass}
\shortauthors{Romani et al.}
\defcitealias{romani2011orbit}{RS11} 	
\defcitealias{Kandel_2020}{KR20}
\defcitealias{2017ApJ...845...42S}{SR17}

\begin{document}

\title{PSR J0952-0607: The Fastest and Heaviest Known Galactic Neutron Star}

\correspondingauthor{R.W. Romani}
\email{rwr@astro.stanford.edu}

\author[0000-0001-6711-3286]{Roger W. Romani}
\affil{Department  of  Physics,  Stanford  University,  Stanford,  CA 94305, USA}
\author[0000-0002-5402-3107]{D. Kandel}
\affil{Department  of  Physics,  Stanford  University,  Stanford,  CA 94305, USA}
\author[0000-0003-3460-0103]{Alexei V. Filippenko}
\affil{Department of Astronomy, University of California, Berkeley, CA 94720-3411, USA}
\author{Thomas G. Brink}
\affil{Department of Astronomy, University of California, Berkeley, CA 94720-3411, USA}
\author{WeiKang Zheng}
\affil{Department of Astronomy, University of California, Berkeley, CA 94720-3411, USA}

\begin{abstract}
We describe Keck-telescope spectrophotometry and imaging of the companion of the ``black widow" pulsar PSR~J0952$-$0607, the fastest known spinning neutron star (NS) in the disk of the Milky Way. The companion is very faint at minimum brightness, presenting observational challenges, but we have measured multicolor light curves and obtained radial velocities over the illuminated ``day" half of the orbit. The model fits indicate system inclination $i=59.8\pm 1.9^\circ$ and a pulsar mass $M_{\rm NS} = 2.35\pm 0.17\, M_\sun$, the largest well-measured mass found to date. Modeling uncertainties are small, since the heating is not extreme; the companion lies well within its Roche lobe and a simple direct-heating model provides the best fit. If the NS started at a typical pulsar birth mass, nearly $1\,M_\odot$ has been accreted; this may be connected with the especially low intrinsic dipole surface field, estimated at $6\times 10^7$\,G. Joined with reanalysis of other black widow and redback pulsars, we find that the minimum value for the maximum NS mass is $M_{\rm max} > 2.19\,M_\odot$\,$(2.09\,M_\odot)$ at $1\sigma$\,$(3\sigma)$ confidence. This is $\sim 0.15\,M_\odot$ heavier than the lower limit on $M_{\rm max}$ implied by the white-dwarf--pulsar binaries measured via radio Shapiro-delay techniques.
\end{abstract}


\section{Introduction} \label{sec:intro}

Pulsar PSR~J0952$-$0607 (hereafter J0952) was discovered by \citet{2017ApJ...846L..20B} with a spin period of $P_s=1.41$\,ms, making it the fastest-spinning pulsar in the disk of the Milky Way. It is a ``black widow" (BW) pulsar with a low-mass (substellar) companion being irradiated and evaporated by the pulsar luminosity. \citet{2019ApJ...883...42N} subsequently detected it as a gamma-ray pulsar, and obtained additional radio timing and optical photometry that allowed initial fits for the system properties. In particular, they measure a low period derivative ${\dot P_{\rm obs}} = 4.6 \times 10^{-21}\,{\rm s\,s^{-1}}$, which places an upper limit on the surface dipole field of $8.2\times 10^7$\,G, among the 15 lowest-known pulsar magnetic fields ($B$), even before the \citet{1970SvA....13..562S} correction. Since in addition their optical photometry suggests a large ($>5$\,kpc) distance, and timing data gave a best-fit (albeit low-significance) proper motion of $\sim 10$\,mas\,yr$^{-1}$, the correction to find the intrinsic ${\dot P}$,
$$
{\dot P}={\dot P}_{\mathrm{obs}} - 2.43\times 10^{-21} P\, \mu_{\rm mas/yr}^2 d_{\mathrm{kpc}}, \eqno(1)
$$
may be substantial, reducing ${\dot P}$ to $2.0\times 10^{-21}\,{\rm s\,s^{-1}}$ and thus $B$ to $B_i \approx 6.1 \times 10^7$\,G. Not all pulsars have a proper motion allowing an intrinsic $B_i$ estimate, but only ten pulsars have a lower $B_i$ in the ATNF catalog \citep[][http://www.atnf.csiro.au/research/pulsar/psrcat]{2005AJ....129.1993M}.

The measurement of high neutron star (NS) masses in some white dwarf (WD)--NS millisecond pulsar binaries, via radio-measured Shapiro delay, has been very influential in driving thinking about the dense-matter equation of state. Two systems have best-estimate masses barely exceeding $2.0\,M_\odot$: J0348+0432 at $2.01\pm 0.04\, M_\odot$ \citep{antoniadis2013massive} and PSR~J0740+6620 at $2.08\pm0.07\,M_\odot$ \citep{2021ApJ...915L..12F}, both $1\sigma$ uncertainties. However, the maximum mass that an NS can reach must depend on its binary evolutionary history. For example, inspection of the double NS--NS binaries alone would lead one to conclude that the typical NS mass is $\sim 1.35\,M_\odot$, with a maximum of $\sim 1.45\,M_\odot$; the millisecond pulsars in these systems are only mildly recycled, with spin periods of tens of ms and moderate magnetic fields. One must therefore examine a variety of pulsar binary classes in a quest to find the most massive NSs.

It has long been argued \citep[e.g.,][]{1974SvA....18..217B, 1990Natur.347..741R} that binary accretion may reduce young pulsar terraGauss fields to millisecond-pulsar values. While the mechanism is unclear, ranging from simple burial to heating-driven ohmic decay \citep[see][for a review]{2017JApA...38...48M}, it seems likely that the amount of mass accretion or its duration play a role in controlling the degree of field reduction. Thus, it is interesting to measure the masses of millisecond pulsars, and the fastest-spinning lowest-field pulsars are good candidates for substantial accretion and high NS mass. Since evolution is driven by angular momentum loss and irradiation bloating, rather than by nuclear evolution, long periods of sub-Eddington rate accretion are expected for the progenitors of ``spider" [BW and ``redback" (RB) millisecond pulsar] binaries before the start of the pulsar-driven evaporation phase; these short-period binaries may host very massive NSs.

\begin{deluxetable}{lccc}[h!!]
\tabletypesize{\footnotesize}
\vspace*{-3mm}
\tablewidth{0pt}
\tablecaption{Keck LRIS Observations\label{tab:obs}}
\tablehead{
\colhead{Session} & \colhead{Spec. Exp. (s)} & \colhead{MJD Range} & \colhead{Phase Range}}
\startdata
\hline
2018A Im& $9\,gi+3\,gr$     & 58455.50562-.57691 &$0.92-1.20$\\
2020A Im& $18\,gi$         & 58932.36032-.47731 &$0.07-0.51$ \\
2018A Spec& $12\times 900$& 58454.52711-.63543 &$0.53-0.94$ \\
2019A Spec& $ 5\times 900$& 58577.39244-.43654 &$0.91-1.08$ \\
2020A Spec& $13\times 600$& 58932.22869-.33973 &$0.61-1.00$ \\
2021B Spec& $21\times 900$& 59610.45450-.62940 &$0.39-1.04$ \\
2022A Spec& $10\times 900$& 59637.27289-.37235 &$0.66-1.03$ \\
2022A2 Spec&$11\times 900$& 59641.26374-.37420 &$0.58-0.99$ \\
 \enddata
 \label{table:obs}
\end{deluxetable}

\vskip -17mm
Thus, J0952 is a particularly attractive candidates for further investigation. However, \citet{2019ApJ...883...42N} state, ``Unfortunately, the optical counterpart of PSR~J0952$-$0607 is too faint ($r\approx 23$ at quadrature when the radial velocity is highest) for spectroscopic radial velocity measurements to be feasible even with 10~m class telescopes.'' This is nearly true, but the exceptional $P_s$ and $B$ inspired our intensive campaign of J0952 Keck LRIS imaging and spectroscopy. Modeling these data, we find that the NS mass is indeed the largest well-measured value to date. Combining this mass with those from new modeling of other spider binaries, we show that these objects put a strong lower bound on the maximum NS mass, appreciably above the values inferred from radio Shapiro delay measurements. This should have deep implications for the dense-matter equation of state.

\section{Observations}

With its $P_b=6.42$\,hr orbital period and faint magnitude, the effective study of J0952 requires substantial amounts of large-telescope time, ideally with superior seeing. Our campaign uses the Keck-I~10\,m telescope and Low Resolution Imaging Spectrometer \citep[LRIS;][]{oke1995keck}. We report here on two dual-color (typically $g/i$ with a few $r/i$ pairs) imaging runs and six spectroscopy campaigns (Table \ref{tab:obs}). For the latter we used the 5600\,\AA\ dichroic, the 600\,l/4000\,\AA\ blue grism, and the 400\,l/8500\,\AA\ red grating, covering $\sim 3300$--10,500\,\AA\ with dispersions of 0.63\,\AA\,pixel$^{-1}$ (blue side) and 1.2\,\AA\,pixel$^{-1}$ (red side). Typical exposures were 600--900\,s. All spectra were processed with the {\tt LPipe} reduction software \citep{2019PASP..131h4503P}. The LRIS red arm thick chip collects many cosmic rays, so additional editing was needed to clean the red-arm spectra. 
All companion and comparison star spectra are available as a DbF (Data behind Figures) tarball.

With the atmospheric dispersion corrector (ADC) being used, we could rotate the $1^{\prime\prime}$-wide slit away from the parallactic angle \citep{Fil82} to simultaneously measure a nearby brighter G1 star with known Pan-STARRS2 (PS2) magnitudes. This allows us to monitor the system throughput and wavelength solution among frames. In particular, since this PS2 star has known and stable magnitudes, we integrate the spectra over the SDSS standard $ugriz$ filter bands using the {\tt sbands} {\tt IRAF} script, and calibrate with the comparison star's catalog magnitudes (converted to the SDSS system using the prescription of \citealt{finkbeiner2016hypercalibration}) to obtain light curves with absolute fluxes, up to small gray shifts from differential slit losses. 

\begin{figure*}[ht!!]
    \centering
    \vspace*{-15mm}\hspace*{-5mm}\includegraphics[scale=0.42]{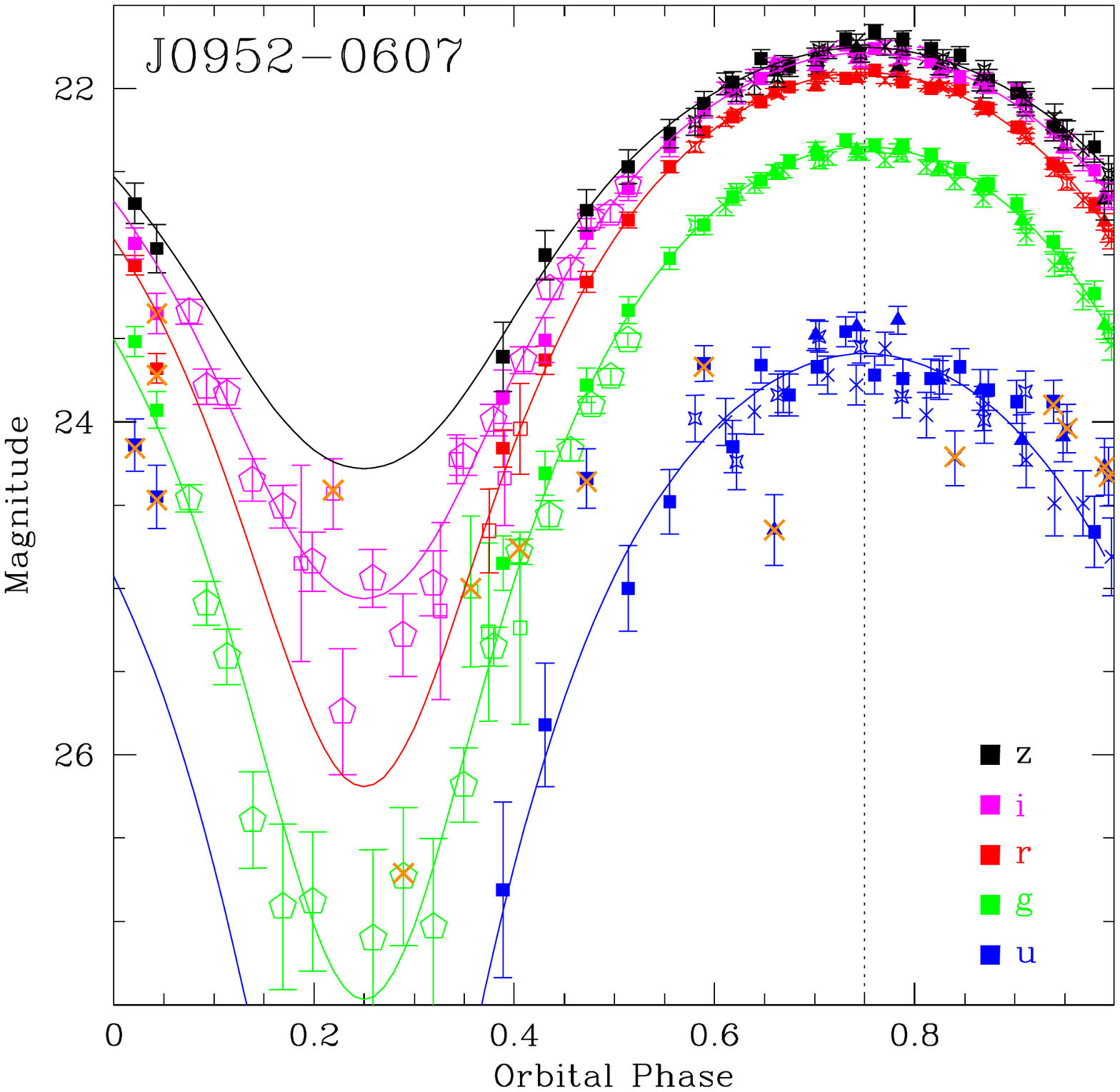}
    \includegraphics[scale=0.42]{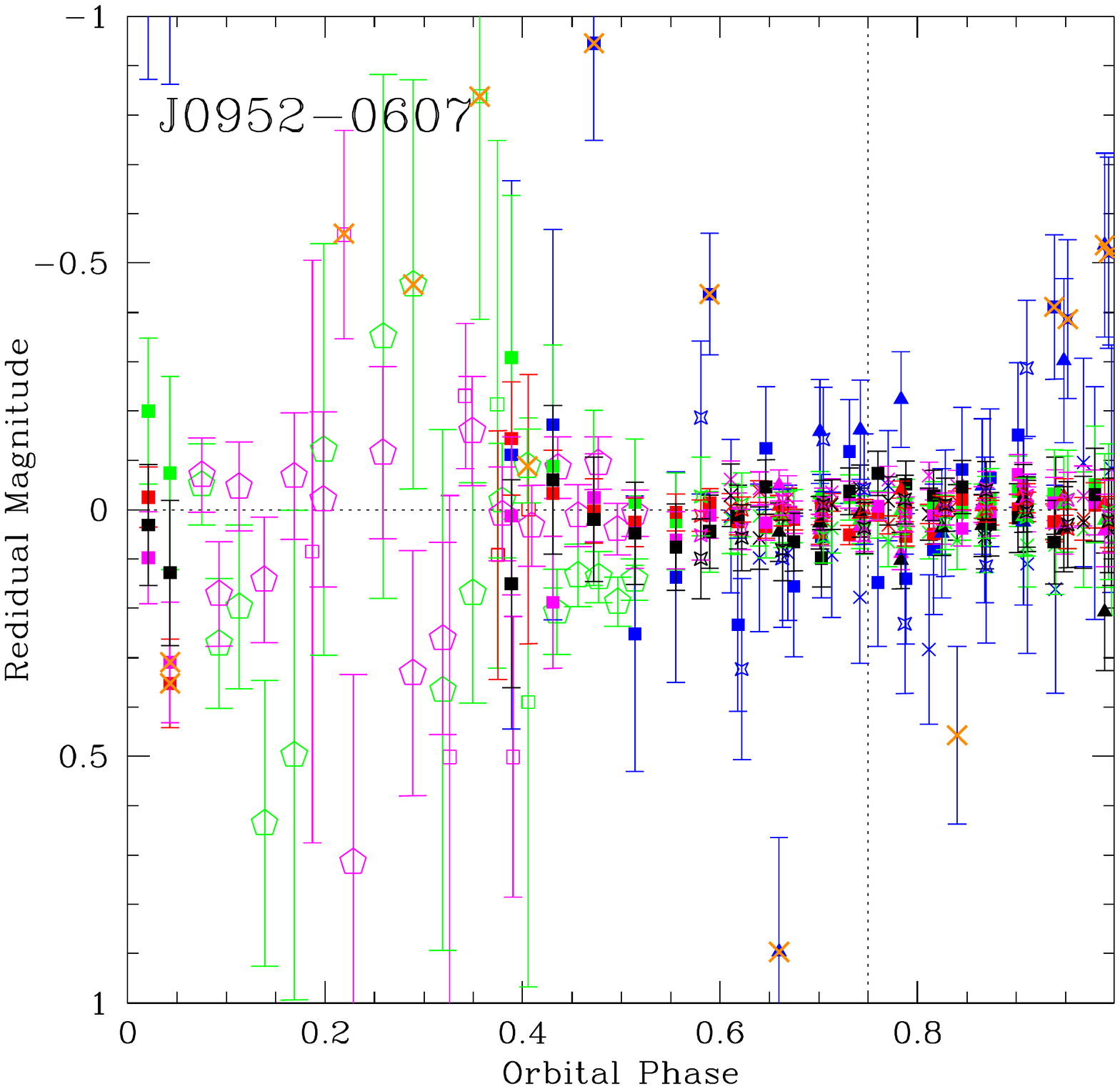}
    \vspace{-25mm}
    \caption{{\it Left:} Calibrated {\it ugriz} light curves from Keck LRIS spectrophotometry (filled dots) and direct imaging (open dots). Outlier points excluded in the ``trimmed" model fitting are indicated by orange crosses. The curves show the best-fit $ugiz$ models and the dotted line marks pulsar inferior conjunction. {\it Right:} Residuals to the best model fit. Optionally excluded points are again marked by orange crosses.} 
    \label{fig:LC}
    \vspace*{-0mm}
\end{figure*}

Our spectroscopic strategy was to observe J0952 for half-nights ideally covering optical maximum brightness. To cover the critical phases at low airmass under dark moon conditions required a careful selection of observing nights. Unfortunately, none of the nights had particularly good seeing; indeed, several additional half-nights requested for this program were completely lost to weather and COVID-19 shutdowns.  Nevertheless, the night-to-night consistency in the flux scale and radial-velocity (RV) scale, aided by the comparison star, has let us make combined fits of all the data, allowing good model constraints.

\begin{figure}[t!]
\centering
    \vspace*{-15mm}\hspace*{-10mm}\includegraphics[scale=0.48]{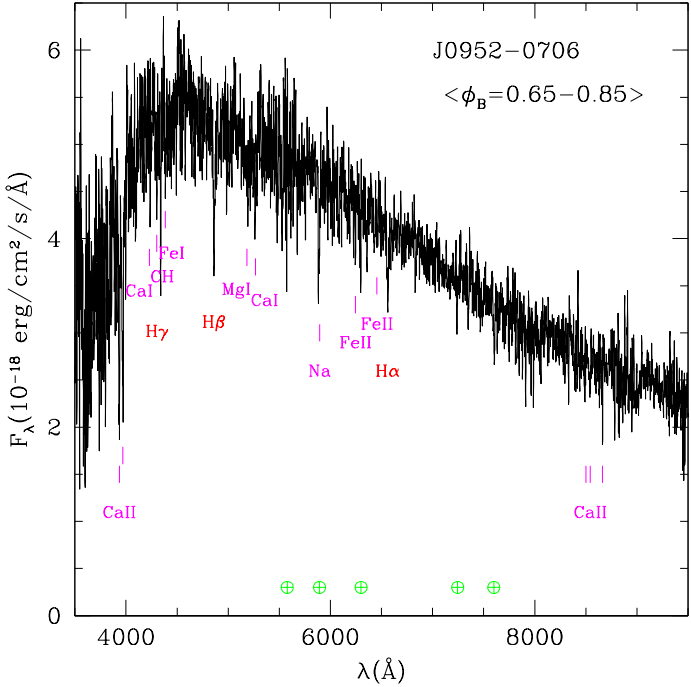}
    \vspace*{-30mm}
\caption{Spectrum of the heated face. S/N-weighted average of 27 individual spectra, Doppler corrected to zero velocity, using the best-fit RV model. Select lines are indicated.}
\label{fig:maxspec}
\end{figure}

As noted, imperfect position angle (PA) and slit placement can give small differential slit losses. These we correct via gray shifts to the {\tt sbands} magnitudes, matching the imaging fluxes in the overlap regions and between epochs. These $\le -0.2$\,mag shifts increase the companion flux; this is sensible, as one tends to ensure that the (brighter) comparison star is well covered by the slit. 

\begin{figure}[t!]
\centering
    \vspace*{-15mm}\hspace*{-10mm}\includegraphics[scale=0.49]{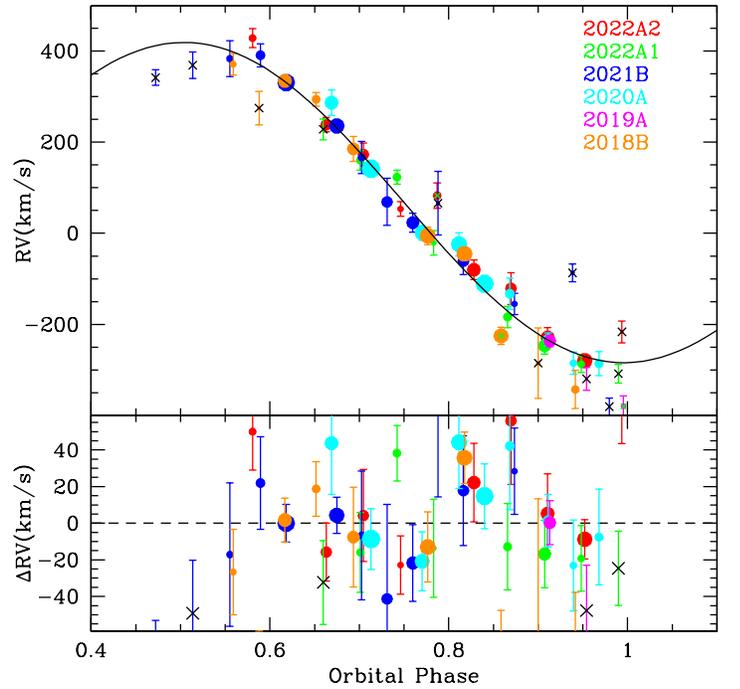}
    \vspace*{-30mm}
\caption{RV measurements for J0952. Dot size is proportional to the correlation coefficient $R$; low-significance correlations with $R<2$ are marked with black crosses and are excluded from all fits. Three higher-significance points marked with green crosses appear as outliers and are optionally excluded for the ``trimmed" fit. The background curve is the best-fit RV curve for the heating model fit to the light curve in Figure 1. The lower panel shows the fit residuals.}
\label{fig:RV}
\end{figure}

In the phase window $\phi_B=0.1$--0.4 the companion is too faint to determine spectrophotometric colors. We therefore collected LRIS dual-band ($g/i$ with a few $g/r$) direct images covering minimum brightness. We established a reference grid of PS2 catalog stars, with positions set relative to the pulsar companion, measured near maximum brightness, and extracted forced aperture photometry at these fixed positions. PS2 magnitudes were converted to the SDSS scale and used to determine the companion magnitude. Direct photometric errors were combined in quadrature with zero-point systematic errors determined from scatter in the catalog star residuals; near minimum the photometric errors strongly dominate the final uncertainty. The resulting light curve covers a large ($>5$\,mag) range, an observational challenge. Image artifacts seem to produce a few modest outliers, and companion flare activity may occasionally brighten the source. We discuss trimming such points below.

We measure the RV amplitude by IRAF {\tt rvsao} \citep{1998PASP..110..934K} cross-correlation with a G1 template, using the 3800--9500\,\AA\ range, excluding wavelengths near strong telluric features. The spectra were resampled to 8192 wavelengths and filtered in Fourier space with a lower cosine-bell filter cutoff from 20--100 and the upper cutoff running from 1000--1800. The average companion spectrum near $\phi_B=0.75$ is shown in Figure \ref{fig:maxspec}. The pulsar and comparison star were correlated using identical parameters. For the first two epochs (2018B and 2019A), a different slit PA led to the use of a cooler ($\sim$ K5) comparison star. A small velocity offset was found between these data and the G1 comparison-star correlations used for the other epochs. In practice, the red K5 star produced poorer RV solutions, and lacked sufficient flux for good $g$ and $u$ calibration; thus, the 2018B and 2019A data do not contribute spectrophotometric points to Figure \ref{fig:LC}. The G1 comparison-star velocity was always measured to $\sigma_v<5$\,km\,s$^{-1}$, but pulsar companion velocity uncertainties varied greatly from as little as 10\,km\,s$^{-1}$ near ``midday'' ($\phi_B=0.75$) to 30--50\,km\,s$^{-1}$ toward the ``dawn' and ``dusk' phases. In addition, spurious correlation peaks appeared well outside the plausible companion-velocity range for many lower signal-to-noise-ratio (S/N) spectra. 

Thus, we measured in two passes, first correlating all spectra over a broad velocity range, and fitting the results with a simple sinusoid RV of amplitude $K$ and offset $\Gamma$ at the radio ephemeris orbital phasing. We then correlated again, accepting only results within $\pm 100$\,km\,s$^{-1}$ of the preliminary fit. This greatly decreased the number of outlier points, and allowed us to recover several weaker RV measurements near the plausible companion solution. For each spectrum the velocity was referenced to that of the comparison star, which varied between exposures, evidently driven by details of the pipeline wavelength solution and extraction aperture. We confirm that these variations are in the pulsar, as well; the scatter in the final RVs increases significantly if the points are not referenced to the comparison star. Small overall RV zero-point shifts were also found in several epochs, again likely owing to differences in the apertures. Figure \ref{fig:RV} shows all RV values with a finite correlation coefficient ($R>0$). The point size is scaled to $R$; a number of obvious outliers with very small $R$ remain.  

\section{Photometric/Radial-Velocity Fitting}

Our light-curve fits are performed with an outgrowth of the {\tt ICARUS} light-curve model \citep{breton2012koi} with the additions described by \citet{Kandel_2020} and \citet{2021ApJ...908L..46R}; the main parameters of interest are the orbital inclination $i$, the companion Roche lobe fill factor $f_1$, the heating luminosity $L_H$, the companion's night-side temperature $T_N$, and the system distance $d$. In addition we can fit for the extinction $A_V$ to the source and a set of ``bandpass calibrations," for possible imperfections in the photometric zero points.

The MCMC fits and Bayseian parameter estimation are performed by Multinest sampling \citep{feroz2009multinest} using its {\tt Python} implementation {\tt pymultinest} \citep{buchner2016pymultinest}.  In addition to the photometry, the fits use the pulsar kinematic parameters of \citet{2019ApJ...883...42N}. The light-curve fit is also (weakly) dependant on the companion center of mass RV $K_{\rm CoM}$. We do not fit for this with {\tt ICARUS}; instead, we initially fit light curves using the observed spectroscopic $K$ and then refit with $K_{\rm CoM}$ determined from the spectroscopic fit (below), iterating to convergence.

The fit $\chi^2$ is increased by a handful of measurements several $\sigma$ from the models. The synthesized $u$ photometry has several such points owing to the very low S/N (per resolution element) of the spectra in this band. The most important outliers are a few bright $g$ and $i$ direct-imaging points during the ``night'' phase. Many spider binaries exhibit occasional short-term ($\sim 1000$\,s) light-curve flares. Since the quiescence light curve is needed for the light-curve fits, we optionally excise the few points well above the model curves (marked in Figure 1). Table 2 gives the fit values with these points first excluded, and then included. Note that inclusion slightly decreases the fit $i$ and increases the mass (by $\sim 0.7\sigma$), while giving much larger $\chi^2$. We therefore adopt the fit to the ``trimmed" dataset as conservative.

\begin{figure}[t!]
\centering
    \vspace*{-1mm}\hspace*{-0mm}\includegraphics[scale=0.3]{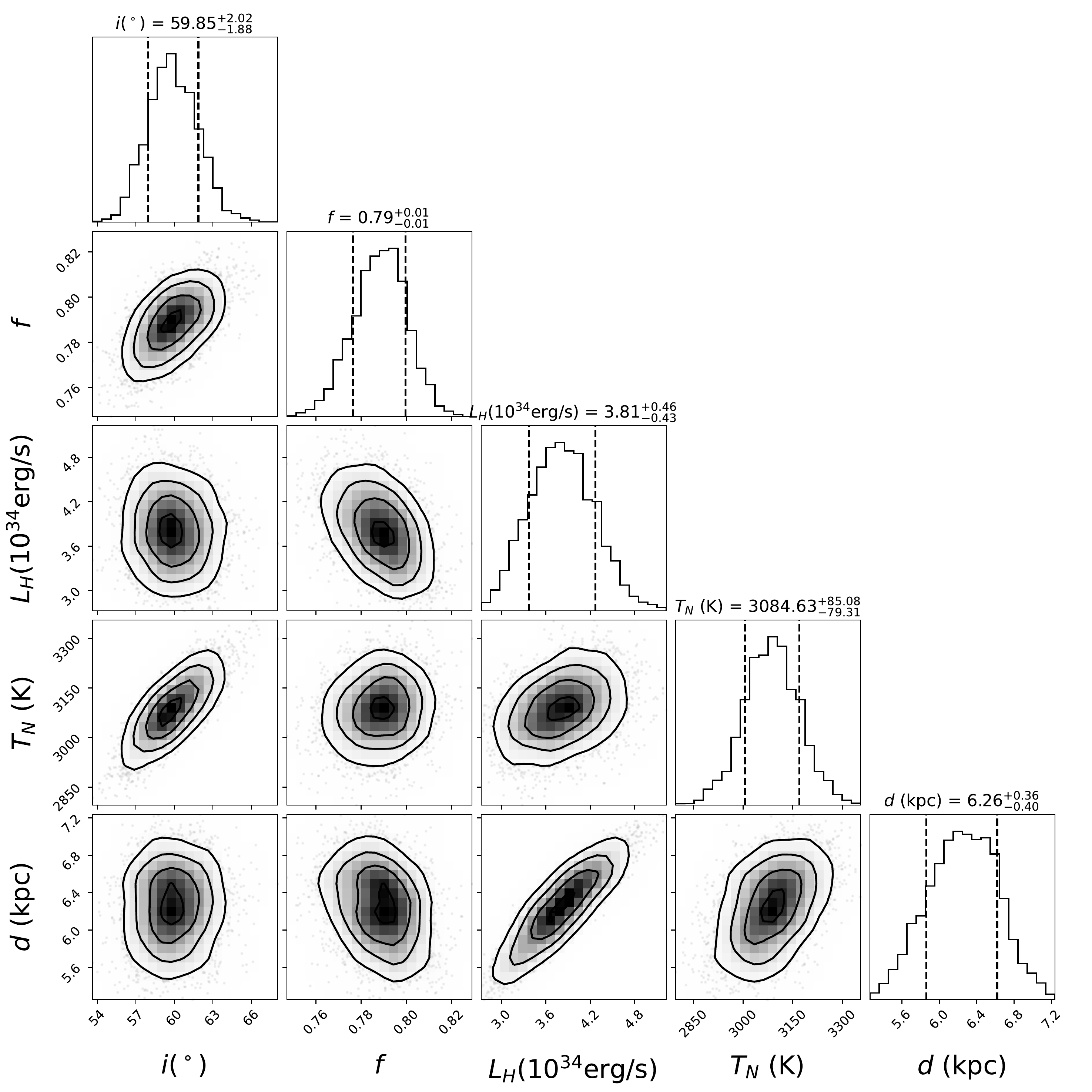}
    \vspace*{-7mm}
\caption{Corner plot from the MCMC light-curve fit to the ``trimmed" photometry points. Contours are at $0.5\sigma$, $1.0\sigma$, $1.5\sigma$, and $2\sigma$.}
\label{fig:corner234}
\end{figure}

Extinction in the direction of J0952 is estimated from the three-dimensional dust map of \citet{2018MNRAS.478..651G}, reaching its maximal $E(g-r)=0.05_{-0.02}^{+0.03}$\,mag ($A_V =0.14^{+0.08}_{-0.05}$\,mag) by 0.3\,kpc. Treating $A_V$ as a free parameter in the MCMC fits, we find $0.16\pm 0.04$\,mag, consistent with the dust-map value. Since we rely on the PS2 $griz$ catalog magnitudes of our comparison star, there are $\sim$0.02\,mag zero point uncertainties \citep[lacking PS2 $u$, we rely on the extrapolation of][]{finkbeiner2016hypercalibration}. Leaving the passband calibrations free in the {\tt ICARUS} fit to the trimmed dataset, we find $\Delta u=+0.07^{+0.08}_{-0.10}$, $\Delta g=+0.05^{+0.05}_{-0.06}$, $\Delta r=+0.02\pm 0.04$, $\Delta i=-0.02\pm 0.04$ and $\Delta z=-0.02\pm 0.04$\,mag, all consistent with zero. We therefore believe that our initial calibration was quite good. For the untrimmed data, the outliers produce a large $\chi^2$/DoF and drive these color terms to larger values, notably $\Delta ugriz=+0.27,\,+0.23,\,+0.03,\,-0.12,\,-0.15$\,mag. While including these passband shifts does not strongly change either $\chi^2$ or the fit parameters, it does slightly increase the MCMC uncertainty ranges of other parameters (most notably $d$ and $L_H$); we thus retain these in the fit, although we omit them from Figure \ref{fig:corner234}.

The fits all find $i \le 60^\circ$, with $f_1=0.79$ (companion underfilling its Roche lobe), $T_N \approx 3000$\,K, day side heated to $\sim 6200\,$K, and $d \approx 6$\,kpc, larger than the dispersion-measure estimates. These values are very similar to those of \citet{2019ApJ...883...42N}, but with substantially higher precision. Interestingly, simple direct-heating models always provide the best statistical fit; adding additional effects required for many other spider binaries, such as companion hot spots and surface winds, does not significantly improve the fit. This means that our results are particularly immune to model-dependant systematics, and the fit-parameter errors (Table \ref{table:fit}) are dominated by measurement uncertainties.

\begin{deluxetable}{lcc}[h!!]
\tabletypesize{\footnotesize}
\tablewidth{0pt}
\tablecaption{Light-Curve/RV-Fit Results for J0952$^a$\label{tab:lc_fit}}
\tablehead{
\colhead{Parameters} & \colhead{Trimmed} & \colhead{All}}
\startdata
$i\, (\mathrm{ deg})$&$59.8^{+2.0}_{-1.9}$&$58.5^{+1.9}_{-1.8}$\\
$f_1$ & $0.79\pm 0.01$ & $0.77\pm 0.01$ \\
$L_{\mathrm{H}}\,/10^{34}\,(\mathrm{erg/s})$ & $3.81^{+0.46}_{-0.43}$  &$6.22^{+0.88}_{-0.77}$\\
$T_{\rm N}$\,(K) & $3085^{+85}_{-80}$ & $3206^{+100}_{-95}$ \\
$d_{\rm kpc}$ & $6.26^{+0.36}_{-0.40}$  & $7.60^{+0.74}_{-0.82}$ \\
 $\chi^2/{\rm DoF}$ & 286/(298-11)[]1.00] &451/(314-11)[1.49] \\
 \hline
 \hline
 $K_{\rm CoM}$\,(km/s) & $376.1\pm 5.1$ & $379.1\pm 6.8$ \\
 $M_{\rm NS}\,(M_\odot)$ & $2.35\pm 0.17$ & $2.50\pm 0.20$  \\
 $M_{\rm C}\,(M_\odot)$ &  $0.032\pm 0.002$ & $0.034\pm 0.002$ \\
 $\chi^2/{\rm DoF}$ &55/(40-2)[1.4]& 90/(43-2)[2.2]\\
 \enddata
 \tablenotetext{}{$^a$Also fitted: $A_V$, $\Delta u$, $\Delta g$, $\Delta r$, $\Delta i$, $\Delta z$. See text.}
\label{table:fit}
\end{deluxetable}
\vskip -15mm

We next compare the Keck spectroscopy with model RVs computed from photometric-fit companion models. As emphasized by \citet{2018ApJ...859...54L} and discussed by \citetalias{Kandel_2020}, different species' temperature sensitivities cause varying line equivalent widths (EWs) across the face of the companion. As the model parameters change, the surface heating changes and the line EWs vary. For J0952, metal lines dominate the cross-correlation signal (as appropriate for our G1 correlation spectrum); we apply metal line equivalent width EW($T$) weighting to correct center-of-mass RVs to observed center-of-light values \citepalias[see][]{Kandel_2020}. If we (inappropriately) apply Balmer EW($T$) weighting, the RV amplitude increases to $384\pm\,5{\rm km\,s^{-1}}$, so again our choice is conservative.

We always exclude the low-significance correlations with $R<2$ in the RV MCMC fits (marked with black crosses in Figure \ref{fig:RV}; note that some crossed points follow the expected curve, but many are quite discrepant). The remaining $\chi^2$ is in fact dominated by three outlier points with small uncertainties (marked with green crosses in Fig. \ref{fig:RV}). In our ``trimmed" fit for the mass we exclude these three points. Again, this is conservative, as with them the $K$ increases slightly (by $0.4\sigma$). The lower section of Table \ref{table:fit} gives the RV fit parameters and their uncertainties. The mass difference in the ``All" column is dominated by the smaller $i$ of the ``All" photometric fit.

\begin{figure}[t!]
\centering
\vspace{-15mm}
\hspace{-5mm}\includegraphics[scale=0.45]{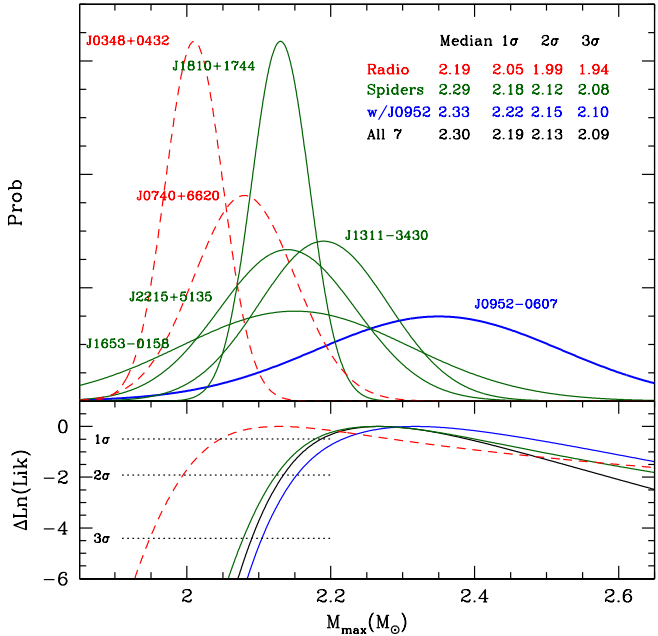}
\vspace{-29mm}
\caption{Mass estimates for heavy NSs. The dashed curves show the two heaviest WD-pulsar binaries with masses measured from radio pulse timing (supplemented by WD atmosphere modeling for J0348). The solid curves show the best mass estimates from companion spectroscopy and light-curve fitting for four BWs (J1810, J1653, J1311 and J0952) and one RB (J2215). 
The bottom panel shows the normalized Ln(Likelihood) for various combinations of these measurements, assuming a flat distribution of masses from 1.8\,$M_\odot$ up to some $M_{\rm max}$. The inset legend gives the median estimator for $M_{\rm max}$ as well as the $1\sigma$, $2\sigma$, and $3\sigma$ lower bounds on its value, for various data subsets.}
\label{fig:masses}
\end{figure}

\section{Mass Implications and Conclusions}

In the quest for precision mass measurements from spider binaries, J0952 has the advantage that the complex modeling effects required to fit other spiders (hot spots, wind flows, extreme gravity darkening) are not needed. This is a natural consequence of its relatively large $P_b$, weak pulsar heating, and low Roche-lobe fill factor. On the other hand, the large distance, weak heating, and low $f_1$ mean that the companion is quite faint, barely bright enough for measurement with present 10\,m telescope systems. Our intensive Keck campaign has managed to measure the binary properties, constraining the masses with sufficient precision that the lower limit on $M_{\rm NS}$ contributes to equation-of-state constraints.

We can examine the minimum required $M_{\rm max}$ by plotting the Gaussian probability distribution functions with ${M_i,\,\sigma_i}$ for observed heavy NSs in Figure \ref{fig:masses}. Here we compare the latest measurements of the $>2\,M_\odot$ ``radio selected" WD--NS MSP J0348+0432 and J0740+6620 with the measurements extracted from light-curve and RV measurements of spider pulsar (BW and RB) companions. For the latter we use the most conservative (best-fit) models with all heating effects as summarized by \cite{kandel2022}. To estimate the minimum value of $M_{\rm max}$ required by these observations, we assume that the NS mass distribution is flat $M^0$ above $M_1=1.8\,M_\odot$ to some cutoff value (for this $>2\,M_\odot$ sample, results are insensitive to the underlying distribution; values change by $<0.01\,M_\odot$ for $M^{-1}$). The log(likelihood) for $n$ measurements is
\begin{equation}
\notag
\begin{split}
{\rm ln}\mathcal{L}= -&n\,{\rm ln}(M_{\rm max}-M_1) \\ 
+&\Sigma_{i=1}^n{\rm ln}\left [{\rm erf} \left( {{M_i-M_1} \over {2^{1/2}\sigma_i}} \right ) -{\rm erf} \left( {{M_i-M_{\rm max}} \over {2^{1/2} \sigma_i}}\right )
\right ].
\end{split}
\end{equation}  
These curves are plotted in the lower panel of Figure \ref{fig:masses}. We list the medians and lower bounds at various confidence levels on $M_{\rm max}$, from the likelihood ratio, for the different sample sets.

Since the spiders are appreciably heavier than the WD--NS binaries, the $1\sigma$ $M_{\rm max}$ bound increases by $\sim 0.13\,M_\odot$, or $\sim 0.17\,M_\odot$ with J0952. We have checked these bounds with several different estimators. For example a bias-corrected, accelerated bootstrap analysis (kindly supplied by B. Efron) of the seven masses gives $M_{\rm max}=2.34\,M_\odot$ with a $2\sigma$ lower bound of $2.17\,M_\odot$. Numerical simulations with random draws and our observed errors give similar values. Of course, J0952 would constrain $M_{\rm max}$ even more strongly if its mass uncertainty could be reduced; with $\sigma=0.05 M_\odot$, we would infer $M_{\rm max}>2.30\,M_\odot \,(1\sigma)$ and $>2.20\, M_\odot\,(3\sigma)$.

From studies of the millisecond pulsar population, it seems that the {\it initial} masses $M_i$ of these pulsars are bimodal, with a dominant component of $M_i \le 1.4\,M_\odot$ and a 20\% subpopulation with $M_i\approx 1.8\,M_\odot$ \citep[][and references therein]{2016arXiv160501665A}. This complicates use of the present mass to constrain mass accretion and test evolutionary models (e.g., of magnetic field reduction). For J0952 at our $1\sigma$ lower limit we infer that at least $0.5\,M_\odot$ and more likely $\sim 1\,M_\odot$ has been accreted. Assuming a start from $1.4\,M_\odot$, three of our BWs have particularly high accreted mass (J0952, $\Delta M \approx 0.95\,M_\odot$; J1653$-$0158, $\Delta M \approx 0.77\,M_\odot$; J1810+1740, $\Delta M \approx 0.73\,M_\odot$). After the \citet{1970SvA....13..562S} corrections, these have some of the lowest pulsar dipole fields known (estimated as $6.1 \times 10^7$\,G, $3.9 \times 10^7$\,G, and $7.7 \times 10^7$\,G, respectively). More $M$ and $B_i$ measurements, as well as additional analysis, are needed to see if this is coincidental or causal. 

Finally, we note that with a central value of $2.35\,M_\odot$, J0952 provides the most severe constraints on the dense-matter equation of state. This remains true even considering the lower bound, and this bound can be adopted with high confidence given the strong preference for a simple heating model. Of course, we would like an even tighter mass measurement of this especially important system. Improved photometry in blue filters at optical minimum may be feasible with 10\,m-class telescopes and excellent conditions, but improved RVs likely await the 30\,m telescope era.

\bigskip

We thank B. Efron for a discussion of bounds from a data sample and for the bootstrap analysis, and the anonymous referee whose probing questions led to some improvements in the paper.
The data presented herein were obtained at the W.~M. Keck Observatory, which is operated as a scientific partnership among the California Institute of Technology, the University of California, and NASA; it was made possible by the generous financial support of the W.~M. Keck Foundation. We are grateful for the excellent assistance of the observatory staff, as well as the Keck Time Allocation Committee and Keck scheduler for accommodating our requests for specific half-nights.
D.K. and R.W.R. were supported in part by NASA grants 80NSSC17K0024 and 80NSSC17K0502. A.V.F.'s group received generous financial assistance from the Christopher R. Redlich Fund, the TABASGO Foundation, and the U.C. Berkeley Miller Institute for Basic Research in Science (where A.V.F. was a Miller Senior Fellow).

\bibliographystyle{aasjournal}
\bibliography{J0952_main}

\begin{thebibliography}{}
\expandafter\ifx\csname natexlab\endcsname\relax\def\natexlab#1{#1}\fi
\providecommand{\url}[1]{\href{#1}{#1}}
\providecommand{\dodoi}[1]{doi:~\href{http://doi.org/#1}{\nolinkurl{#1}}}
\providecommand{\doeprint}[1]{\href{http://ascl.net/#1}{\nolinkurl{http://ascl.net/#1}}}
\providecommand{\doarXiv}[1]{\href{https://arxiv.org/abs/#1}{\nolinkurl{https://arxiv.org/abs/#1}}}

\bibitem[{{Antoniadis} {et~al.}(2016){Antoniadis}, {Tauris}, {Ozel}, {Barr},
  {Champion}, \& {Freire}}]{2016arXiv160501665A}
{Antoniadis}, J., {Tauris}, T.~M., {Ozel}, F., {et~al.} 2016, arXiv e-prints,
  arXiv:1605.01665.
\newblock \doarXiv{1605.01665}

\bibitem[{Antoniadis {et~al.}(2013)Antoniadis, Freire, Wex, Tauris, Lynch, van
  Kerkwijk, Kramer, G, Dhillon, Driebe, {et~al.}}]{antoniadis2013massive}
Antoniadis, J., Freire, P.~C., Wex, N., {et~al.} 2013, Science, 340

\bibitem[{{Bassa} {et~al.}(2017){Bassa}, {Pleunis}, {Hessels}, {Ferrara},
  {Breton}, {Gusinskaia}, {Kondratiev}, {Sanidas}, {Nieder}, {Clark}, {Li},
  {van Amesfoort}, {Burnett}, {Camilo}, {Michelson}, {Ransom}, {Ray}, \&
  {Wood}}]{2017ApJ...846L..20B}
{Bassa}, C.~G., {Pleunis}, Z., {Hessels}, J.~W.~T., {et~al.} 2017, \apjl, 846,
  L20, \dodoi{10.3847/2041-8213/aa8400}

\bibitem[{{Bisnovatyi-Kogan} \& {Komberg}(1974)}]{1974SvA....18..217B}
{Bisnovatyi-Kogan}, G.~S., \& {Komberg}, B.~V. 1974, \sovast, 18, 217

\bibitem[{Breton {et~al.}(2012)Breton, Rappaport, van Kerkwijk, \&
  Carter}]{breton2012koi}
Breton, R.~P., Rappaport, S.~A., van Kerkwijk, M.~H., \& Carter, J.~A. 2012,
  \apj, 748, 115

\bibitem[{Buchner(2016)}]{buchner2016pymultinest}
Buchner, J. 2016, Astrophysics Source Code Library, ascl

\bibitem[{Feroz {et~al.}(2009)Feroz, Hobson, \& Bridges}]{feroz2009multinest}
Feroz, F., Hobson, M., \& Bridges, M. 2009, Monthly Notices of the Royal
  Astronomical Society, 398, 1601

\bibitem[{Filippenko(1982)}]{Fil82}
Filippenko, A.~V. 1982, \pasp, 94, 715

\bibitem[{Finkbeiner {et~al.}(2016)Finkbeiner, Schlafly, Schlegel, Padmanabhan,
  Juri{\'c}, Burgett, Chambers, Denneau, Draper, Flewelling,
  {et~al.}}]{finkbeiner2016hypercalibration}
Finkbeiner, D.~P., Schlafly, E.~F., Schlegel, D.~J., {et~al.} 2016, \apj, 822,
  66

\bibitem[{{Fonseca} {et~al.}(2021){Fonseca}, {Cromartie}, {Pennucci}, {Ray},
  {Kirichenko}, {Ransom}, {Demorest}, {Stairs}, {Arzoumanian}, {Guillemot},
  {Parthasarathy}, {Kerr}, {Cognard}, {Baker}, {Blumer}, {Brook}, {DeCesar},
  {Dolch}, {Dong}, {Ferrara}, {Fiore}, {Garver-Daniels}, {Good}, {Jennings},
  {Jones}, {Kaspi}, {Lam}, {Lorimer}, {Luo}, {McEwen}, {McKee}, {McLaughlin},
  {McMann}, {Meyers}, {Naidu}, {Ng}, {Nice}, {Pol}, {Radovan},
  {Shapiro-Albert}, {Tan}, {Tendulkar}, {Swiggum}, {Wahl}, \&
  {Zhu}}]{2021ApJ...915L..12F}
{Fonseca}, E., {Cromartie}, H.~T., {Pennucci}, T.~T., {et~al.} 2021, \apjl,
  915, L12, \dodoi{10.3847/2041-8213/ac03b8}

\bibitem[{{Green} {et~al.}(2018){Green}, {Schlafly}, {Finkbeiner}, {Rix},
  {Martin}, {Burgett}, {Draper}, {Flewelling}, {Hodapp}, {Kaiser}, {Kudritzki},
  {Magnier}, {Metcalfe}, {Tonry}, {Wainscoat}, \&
  {Waters}}]{2018MNRAS.478..651G}
{Green}, G.~M., {Schlafly}, E.~F., {Finkbeiner}, D., {et~al.} 2018, \mnras,
  478, 651, \dodoi{10.1093/mnras/sty1008}

\bibitem[{Kandel \& Romani(2020)}]{Kandel_2020}
Kandel, D., \& Romani, R.~W. 2020, \apj, 892, 101,
  \dodoi{10.3847/1538-4357/ab7b62}

\bibitem[{Kandel \& Romani(2022)}]{kandel2022}
---. 2022, \apj, in prep

\bibitem[{{Kurtz} \& {Mink}(1998)}]{1998PASP..110..934K}
{Kurtz}, M.~J., \& {Mink}, D.~J. 1998, \pasp, 110, 934, \dodoi{10.1086/316207}

\bibitem[{{Linares} {et~al.}(2018){Linares}, {Shahbaz}, \&
  {Casares}}]{2018ApJ...859...54L}
{Linares}, M., {Shahbaz}, T., \& {Casares}, J. 2018, \apj, 859, 54,
  \dodoi{10.3847/1538-4357/aabde6}

\bibitem[{{Manchester} {et~al.}(2005){Manchester}, {Hobbs}, {Teoh}, \&
  {Hobbs}}]{2005AJ....129.1993M}
{Manchester}, R.~N., {Hobbs}, G.~B., {Teoh}, A., \& {Hobbs}, M. 2005, \aj, 129,
  1993, \dodoi{10.1086/428488}

\bibitem[{{Mukherjee}(2017)}]{2017JApA...38...48M}
{Mukherjee}, D. 2017, Journal of Astrophysics and Astronomy, 38, 48,
  \dodoi{10.1007/s12036-017-9465-6}

\bibitem[{{Nieder} {et~al.}(2019){Nieder}, {Clark}, {Bassa}, {Wu}, {Singh},
  {Donner}, {Allen}, {Breton}, {Dhillon}, {Eggenstein}, {Hessels}, {Kennedy},
  {Kerr}, {Littlefair}, {Marsh}, {Mata S{\'a}nchez}, {Papa}, {Ray}, {Steltner},
  \& {Verbiest}}]{2019ApJ...883...42N}
{Nieder}, L., {Clark}, C.~J., {Bassa}, C.~G., {et~al.} 2019, \apj, 883, 42,
  \dodoi{10.3847/1538-4357/ab357e}

\bibitem[{Oke {et~al.}(1995)Oke, Cohen, Carr, Cromer, Dingizian, Harris,
  Labrecque, Lucinio, Schaal, Epps, {et~al.}}]{oke1995keck}
Oke, J., Cohen, J., Carr, M., {et~al.} 1995, Publications of the Astronomical
  Society of the Pacific, 107, 375

\bibitem[{{Perley}(2019)}]{2019PASP..131h4503P}
{Perley}, D.~A. 2019, \pasp, 131, 084503, \dodoi{10.1088/1538-3873/ab215d}

\bibitem[{{Romani}(1990)}]{1990Natur.347..741R}
{Romani}, R.~W. 1990, \nat, 347, 741, \dodoi{10.1038/347741a0}

\bibitem[{{Romani} {et~al.}(2021){Romani}, {Kandel}, {Filippenko}, {Brink}, \&
  {Zheng}}]{2021ApJ...908L..46R}
{Romani}, R.~W., {Kandel}, D., {Filippenko}, A.~V., {Brink}, T.~G., \& {Zheng},
  W. 2021, \apjl, 908, L46, \dodoi{10.3847/2041-8213/abe2b4}

\bibitem[{{Shklovskii}(1970)}]{1970SvA....13..562S}
{Shklovskii}, I.~S. 1970, \sovast, 13, 562

\end{thebibliography}
\end{document}